\documentclass[prd,aps,preprintnumbers,nofootinbib,twocolumn]{revtex4}

\usepackage{amscd}
\usepackage{amsmath, bbm} 
\usepackage{hyperref}
\usepackage{graphicx}
\usepackage{lipsum}
\usepackage{color}
\usepackage{url}

\newcommand{\be}{\begin{equation}}
\newcommand{\ee}{\end{equation}}
\newcommand{\bea}{\begin{eqnarray}}
\newcommand{\eea}{\end{eqnarray}}
\newcommand{\beq}{\begin{equation}}
\newcommand{\eeq}{\end{equation}}

\usepackage{tikz} 
\usetikzlibrary{shapes,arrows,positioning,automata,backgrounds,calc,er,patterns}
\usepackage{tikz-feynman}

\begin{document}
\preprint{CERN-TH-2020-113}

\title{Undulating Dark Matter}

\author{Joe Davighi} \email{jed60@cam.ac.uk} \affiliation{DAMTP, University of Cambridge, Wilberforce Road, Cambridge, UK}

\author{Matthew McCullough} \email{matthew.mccullough@cern.ch} \affiliation{CERN, Theoretical Physics Department, Geneva, Switzerland} \affiliation{DAMTP, University of Cambridge, Wilberforce Road, Cambridge, UK}

\author{Joseph Tooby-Smith} \email{jss85@cam.ac.uk} \affiliation{Cavendish Laboratory, University of Cambridge, Cambridge, UK}

\begin{abstract}
We suggest that an interplay between microscopic and macroscopic physics can give rise to dark matter (DM) whose  interactions with the visible sector fundamentally undulate in time, independent of celestial dynamics. A concrete example is provided by fermionic DM with an electric dipole moment (EDM) sourced by an oscillating axion-like field, resulting in undulations in the scattering rate.  The discovery potential of light DM searches can be enhanced by additionally searching for undulating scattering rates, especially in detection regions where background rates are large and difficult to estimate, such as for DM masses in the vicinity of 1 MeV where DM-electron scattering dominantly populates the single electron bin.  An undulating signal could also reveal precious dark sector information after discovery.  In this regard we emphasise that, if the recent XENON1T excess of events is due to light DM scattering exothermically off electrons, future analyses of the time-dependence of events could offer clues as to the microscopic origins of the putative signal.
\end{abstract}

\maketitle

\section{Introduction}
\label{sec:intro}
The evidence for dark matter is unequivocal, yet our ignorance of the dark sector remains as vast as the Universe it shapes.  The experimental effort to detect it is proportionately extensive, pushing back frontiers on mass scales ranging from cosmologically light bosonic fields to astronomically massive objects.  Direct detection experiments aimed at observing the scattering of galactic dark matter particles on matter within the laboratory have formed a significant component of this global effort. This strategy has for decades focussed on searching for weakly interacting massive particles (WIMPs) scattering off nuclei.  However, in the last decade, due to a growing synergy between advancing experimental techniques
and an evolving theoretical landscape, a new frontier below the GeV scale is opening, with a burgeoning number of experiments and analysis techniques planned and proposed.

As a smoking gun for galactic dark matter scattering and also as a powerful background mitigation strategy, the modulation of dark matter scattering rates on diurnal and annual timescales has been a useful tool in direct detection strategies.  Both timescales follow from the time-dependent angle of attack of the detector into the prevailing galactic dark matter wind.  Even though the precise form of these modulations depends on the details of the local dark matter velocity distribution and on the dark matter scattering process (see {\em e.g.}~\cite{Lee:2015qva} for the case of dark matter-electron scattering, which is most relevant to this paper), the existence of some modulating component on these timescales is certain.  It is then natural to ask whether we should expect solar system or dark matter dynamics to be the sole origin of a modulation? In particular, might there be a fundamental microscopic mechanism that modulates the dark matter scattering rate?
To that end, we here propose a model of dark matter interactions where the coupling between the visible and dark sectors, and therefore the dark matter scattering cross-section, undulates in time with a period which is unrelated to celestial dynamics. We refer to this class of models as Undulating Dark Matter (UDM). 

We expect that looking for an undulation in the dark matter scattering rate would be an especially useful aide in searches for sub-GeV mass dark matter, for example by dark matter-electron scattering. Here, background rates in the single electron bin are typically high and challenging to model confidently, in contrast to the multi-electron bins. Even if the background rate were only poorly understood, we find that an undulating contribution to the signal can be constrained several orders of magnitude better than a constant contribution, for a reasonably wide window of frequencies.\footnote{For a given experiment, the window of frequencies in which we can hunt for an undulation lies between by the inverse total exposure time of the experiment, and the rate at which the data are read out.} This strategy could boost discovery potential in high background search regions, such as the single-electron event bins of the upcoming SENSEI~\cite{Tiffenberg:2017aac,Crisler:2018gci,2004_11378} and SuperCDMS experiments~\cite{Agnese:2016cpb,Agnese:2018col}, and the electronic recoil mode of the XENON1T experiment~\cite{Aprile:2017aty} (in which an excess of events has recently been observed~\cite{Aprile:2020tmw}).

In the particular model that we propose, which should be thought of as just one possible mechanism for UDM, the dark matter interacts with the visible sector via an undulating electric dipole moment (UEDM).
Models in which DM interacts with the visible sector predominantly via an EDM operator comprise a viable and phenomenologically relevant class of scenarios which have been studied for some time \cite{Pospelov:2000bq,Sigurdson:2004zp,Ho:2012bg,Schmidt:2012yg,Kopp:2014tsa,Ibarra:2015fqa,Sandick:2016zut,Kavanagh:2018xeh,Chu:2018qrm}.  Since it violates $CP$, the discovery of a DM EDM would carry with it not only the triumph of discovery, but also precious microscopic information about the dark sector. 

A sizeable EDM would point towards a microscopic marriage between electromagnetic substructure and $CP$ violation.  The substructure of a neutral DM particle may be confined, as it is for the neutron ({\em e.g.}~\cite{Banks:2010eh,Bagnasco:1993st}), or it may be perturbative, simply implying some perturbative coupling of the dark matter to additional massive charged states ({\em e.g.}~\cite{Graham:2012su,Chang:2019xva}).  In the latter case these states may be within or beyond the Standard Model, but they should not be too decoupled if we want the DM EDM to be observable.  Regarding $CP$ violation, in either case we may take our cue from nature.

It is natural to expect that if we have electromagnetic substructure in tandem with order-one $CP$ violation then we should expect order-one EDMs, in whatever the natural units are. Yet, for QCD, measurements contradict this expectation to an absurd degree, requiring the EDM to be many orders of magnitude below the na\"ive expectation. Unexpected fine-tuning of parameters, the Nelson--Barr mechanism~\cite{Nelson:1983zb,Barr:1984qx,Barr:1984fh}, and the axion~\cite{Peccei:1977hh,Peccei:1977ur,Weinberg:1977ma,Wilczek:1977pj} all offer possible answers to this puzzle.  If the latter were the option preferred by nature,  then it is tempting to speculate that whenever an EDM is connected to an underlying anomalous symmetry in the UV, nature may deliver an axion to relax it in the IR, even in the dark sector. Indeed, looking rather deep into the UV, 
additional light scalars tend to abound in the low-energy limit of string theory (see {\em e.g.}~\cite{Svrcek:2006yi,Arvanitaki:2009fg}).

With these considerations it seems plausible that, in scenarios where DM may possess an EDM, it may also be relaxed by a hidden axion-like particle (ALP).  While superficially this may appear to suppress the discovery opportunity, it also presents a new opportunity if the ALP is cosmologically abundant, since the dark EDM would then undulate at a frequency set by the mass of the ALP.  If the dark EDM expectation value mediates some significant fraction of the dark matter scattering (at least in some region of the parameter space) then this scattering cross-section would itself undulate, giving rise to UEDM signatures that could boost detection prospects.

In Sec.~\ref{sec:EFT} we will consider this scenario from an effective field theory (EFT) perspective, postponing as many microscopic questions as possible to the UV, while retaining the model-independent predictions for the IR. For this, we sketch some possible UV scenarios in Appendix \ref{app}. In Sec.~\ref{sec:cosmology} we describe the cosmology of the ALP to explore the amplitude and frequency of ALP oscillations possible today.  In Sec.~\ref{sec:dd}, guided by the model of UEDM, we come to study the phenomenology and future prospects for the direct detection of undulating dark matter, before concluding.

\section{An effective field theory description}
\label{sec:EFT}

We consider extending the Standard Model (SM) by a stable neutral Dirac fermion $\chi$ with mass $m_\chi$, which we suppose constitutes the bulk of the dark matter, as well as a light pseudo-scalar dark axion field $\phi$, with mass $m_\phi$. Both are SM singlets. The dark axion could arise as the would-be Goldstone boson associated to the breaking of an anomalous global $U(1)_A$ symmetry at some high energy scale $f$, \`a la Peccei and Quinn (PQ)~\cite{Peccei:1977hh,Peccei:1977ur}. 

It is reasonable to suppose that this pair of dark (or, perhaps more accurately, `faint') particles interacts with the SM predominantly via electric and magnetic dipole moment (EDM and MDM respectively) couplings to the photon. When ordered by increasing powers of the $CP$-odd field $\phi$, which we take to be the only source of $CP$ violation in the dark sector, the leading interaction that involves both $\chi$ and $\phi$ comes from the couplings
\be
\mathcal{L}_{\text{DM}} = \mathcal{L}_{\chi} + \mathcal{L}_{\phi} - c \frac{e}{2 \Lambda} \sin \left(\frac{\phi}{f} \right) \overline{\chi} \sigma^{\mu\nu} i\gamma_5 \chi \; F_{\mu\nu} +\text{MDM}~~, \label{eq:lagrangian}
\ee
where $e$ is the electron charge, $\Lambda$ is the scale associated with the dark sector dynamics, and $c$ is a model-dependent dimensionless parameter. We emphasize that the MDM couplings, being $CP$-conserving,  must be even under $\phi \to -\phi$ and so their $\phi$-dependence begins at order $(\phi/f)^2$, a contribution that we do not write explicitly and will henceforth neglect.\footnote{In the absence of the pseudoscalar $\phi$, other sources of $CP$-violation must be present in order for $\chi$ to acquire an EDM. For example, if the DM were a dark baryon, the presence of a $CP$-violating phase in dark quark mixing (analogous to the $CP$-violating CKM phase in the SM) would lead to $CP$-violating four-Fermion operators, themselves suppressed by some factor $\sim 1/\Lambda^2_{CP}$ where $\Lambda_{CP} > \Lambda$. In this scenario, one would expect the effective EDM to scale like $\sim \Lambda/\Lambda^2_{CP}$~\cite{Bagnasco:1993st,Banks:2010eh}. In our model, we emphasize that the $CP$-violating parameter is simply the pseudoscalar $\langle\phi\rangle$, and the correct $CP$-suppression is automatically encoded in the powers of $\phi/f$.}

The coupling $c$, as well as the relation between the scale $\Lambda$ and the dark matter mass $m_\chi$, depend on the microscopic details underlying this effective theory, which therefore sets the scale at which we expect to observe undulating dark matter EDMs. If the dark sector is strongly-coupled, for example, we would expect $c$ to be order-one, while in the weak-coupling r\'egime we would expect $c$ to be suppressed by a loop factor. In Appendix \ref{app}, we comment on the ballpark magnitude of the overall Wilson coefficient 
\be
d_\chi \equiv ce/\Lambda
\ee
in three simple possibilities for the underlying dark sector microscopics:
\begin{enumerate}
\item Strictly QCD-like dark sector dynamics, in which the DM is a neutral baryon akin to the neutron;
\item The DM is a neutral baryon of a QCD-like theory but where the number of colours $N_c$ is very large;
\item The DM is an elementary fermion whose EDM arises perturbatively.
\end{enumerate}
The point of these UV considerations is to highlight how the expected size of the Wilson coefficient $d_\chi$ can depend strongly on the microscropic details of the theory. When we come to study UEDMs in direct detection in \S \ref{sec:dd}, we will briefly discuss implications for the type of microscopic theory we might expect to underlie the EFT (\ref{eq:lagrangian}).

\begin{figure*}[t]
\begin{center}
\begin{tikzpicture}
\begin{feynman}
\vertex (a) {\footnotesize $\phi$};
\vertex [right=1in of a] (b);
\vertex [right=1in of b] (c) {\footnotesize $\phi$};
\vertex [above=of b] (d);
\node at (b) [circle,fill,inner sep=2pt,label=below:{\footnotesize $d_\chi/f^2$}]{};
\node at (d) [circle,fill,inner sep=2pt,label=above:{\footnotesize $d_\chi$}]{};
 
\vertex [right=0.8in of c] (te);

\vertex [above=0.2in of te] (a2) {\footnotesize $\phi$};
\vertex [right=0.9in of a2] (b2);
\vertex [right=0.7in of b2] (c2) ;
\vertex [right=0.9in of c2] (d2) { \footnotesize $\phi$};
\node at (b2) [circle,fill,inner sep=2pt,label={[xshift=-0.4cm, yshift=-0.8cm]\footnotesize$d_\chi/f$}]{};
\node at (c2) [circle,fill,inner sep=2pt,label={[xshift=0.4cm, yshift=-0.8cm]\footnotesize$d_\chi/f$}]{};
\diagram* {
(a) -- [scalar] (b)  -- [scalar] (c),
(b) -- [photon,edge label={\footnotesize $\gamma$}] (d),
(b) -- [solid, half left,edge label={\footnotesize $\chi$}] (d),
(b) -- [solid, half right,edge label={\footnotesize $\chi$},swap] (d),

(a2) -- [scalar] (b2),
(c2) -- [scalar] (d2),
(b2) -- [photon,edge label={\footnotesize $\gamma$}] (c2),
(b2) -- [solid, half left,edge label={\footnotesize $\chi$}] (c2),
(b2) -- [solid, half right,edge label={\footnotesize $\chi$},swap] (c2),
};
\end{feynman}
\end{tikzpicture}
\end{center}
\caption{Two loop diagrams which give the schematic scaling of corrections to $m_\phi^2$ within the EFT described by (\ref{eq:lagrangian}). {\em Left}: magnetic contribution (for which both vertices feature an even number of $\phi$ insertions). {\em Right}: electric contribution (vertices feature an odd number of $\phi$ insertions). Both diagrams lead to mass corrections of the same order, as given by Eq. (\ref{eq:EFTbound}). \label{fig:2loops}}    
\end{figure*}
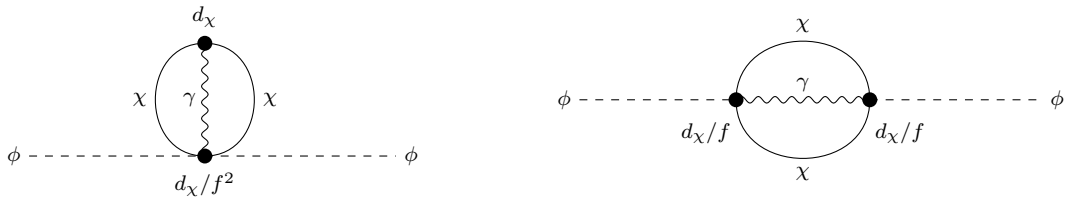

Until then, we will take $d_\chi$ as a purely phenomenological parameter which dictates the low-energy physics. Nonetheless, even in the context of the EFT we cannot take $d_\chi$ to be arbitrarily large for any value of $m_\chi$ and $m_\phi$, because the interactions in Eq.~(\ref{eq:lagrangian}) allow corrections to both the fermion and dark axion mass, dependent on the microscopics. As we will soon see, the mass of the dark axion sets the frequency of the undulating EDM signal. While this frequency is {\em a priori} a free parameter, we will be most interested in cases where it is of the order of inverse months to minutes, corresponding to $10^{-20} \text{~eV} \lesssim m_\phi \lesssim 10^{-15}$ eV or so. Because of this the dark axion must be far lighter than the fermion, and the stronger bound on $d_\chi$ comes from the expected correction to $m_\phi$. Based on $\hbar$ counting we estimate 
\be
\delta m_\phi^2 \sim \frac{1}{4} \left(\frac{1}{16\pi^2}\right)^2 \frac{d_\chi^2 m_\chi^6}{f^2}~~,
\ee
motivated by the cutoff dependence of the two-loop diagrams shown in Fig.~\ref{fig:2loops}, in which we assume the cutoff scale in the loop diagrams is of order $m_\chi$, and have also considered the MDM contributions. Consequently, requiring any such corrections to be below the dark axion mass, we limit
\be
d_\chi \lesssim 32 \pi^2  \frac{m_\phi f}{m^3_\chi} ~~. \label{eq:EFTbound}
\ee
We emphasize that this rough `bound', while imprecise up to order-one factors, is a model-independent estimate of the upper limit on the Wilson coefficient $d_\chi$ that follows from self-consistency of the couplings in the EFT alone in the absence of any additional protection mechanism, such as low-scale supersymmetry in the dark sector. In specific microscopic descriptions, such as the three listed above (and discussed in Appendix \ref{app}), $d_\chi$ will often be required to be smaller still.

\section{Undulating dark EDMs} \label{sec:cosmology}
The lagrangian terms in Eq. \ref{eq:lagrangian} give rise to an effective EDM $d^\text{eff}_\chi(t)$ for the dark matter which, due to $\phi$, is time-dependent, where
\be \label{eq:dchi_eff}
d^\text{eff}_\chi(t) \equiv d_\chi \sin \left(\frac{\phi(t)}{f} \right) ~~.
\ee
Supposing the (light) dark axion field $\phi$ is cosmologically present, its amplitude will be well-approximated by solving the classical equations of motion. In an FLRW background cosmology, one has 
\be
\ddot \phi+3H\dot \phi+m_\phi^2\phi=0 ~~.
\ee
Assuming $m_\phi^2\phi \gg 3H\dot \phi$, the dark axion undergoes weakly damped oscillations. On short enough time scales (with respect to $H^{-1}(t_0)$),\footnote{In fact, things are a little more subtle, because the virialization of any light scalar dark matter particle in the galactic halo leads to an effective quality factor of $10^6$~\cite{Berlin:2019ahk} (see also {\em e.g.}~\cite{Sikivie:2009qn,Davidson:2014hfa} for further discussions of the coherence properties of a cosmologically-present axion-like particle), meaning that the assumption of harmonic oscillations is only appropriate on timescales shorter than $\tau \sim 10^6/m_\phi$. In the contexts of the direct detection experiments we consider in \S \ref{sec:dd}, this is a safe assumption. } we can assume that $\phi(t)$ oscillates harmonically with frequency $m_\phi$.

The amplitude of these oscillations, call it $\phi_0$, is important to the phenomenology of our dark sector, since it sets the size of the effective dark matter EDM. If the dark axion constitutes some fraction $r$ of the local dark matter energy density $\rho_{\text{DM}} \approx 0.3 \text{~GeV}/\text{cm}^{3}$, then
\be \label{eq:phi0}
\frac{\phi_0}{f} \approx \sqrt{r}\; \frac{2 \times 10^{-15} \text{~MeV}^2}{m_\phi f} ~~.
\ee
We will suppose that the dark fermion $\chi$ (rather than the dark axion) constitutes most of the dark matter, so the reader can keep in mind a value of $r\lesssim 0.1$.

Expanding Eq.~\ref{eq:dchi_eff} to leading order in $\phi/f$ and substituting in Eq.~\ref{eq:phi0}, one expects $d^\text{eff}_\chi(t) \approx |d^\text{eff}_\chi| \cos (m_\phi t)$, with the modulation frequency equal to the mass of the dark axion, and with amplitude $ |d^\text{eff}_\chi|\approx d_\chi \phi_0/f$, which can be expressed as
\be \label{eq:modulatingDM}
\frac{|d^\text{eff}_\chi|}{\mu_B} \approx \frac{c}{\Lambda} \frac{\sqrt{r}}{m_\phi f}\; 2\times 10^{-15} \text{~MeV}^3 ~~,
\ee
in units of the Bohr magneton $\mu_B = e/2m_e$. The EFT bound (\ref{eq:EFTbound}) implies an upper limit\be \label{eq:finetuningconst}
\frac{|d^\text{eff}_\chi|}{\mu_B} \lesssim 2 \times 10^{-12} \left(\frac{\text{MeV}}{m_\chi}\right)^3 \sqrt{r} ~~,
\ee
which turns out to be independent of $f$ and $m_\phi$, depending only on the dark matter mass $m_\chi$. The line on which this bound is saturated is included (in blue) in Fig.~\ref{fig:EDMconstraints}.


\section{Detecting undulating dark matter} \label{sec:dd}

Now we discuss the detection prospects, in rather general terms.

\subsection*{Existing constraints}

\begin{figure*}[t]
\begin{center}
\includegraphics[width=4in]{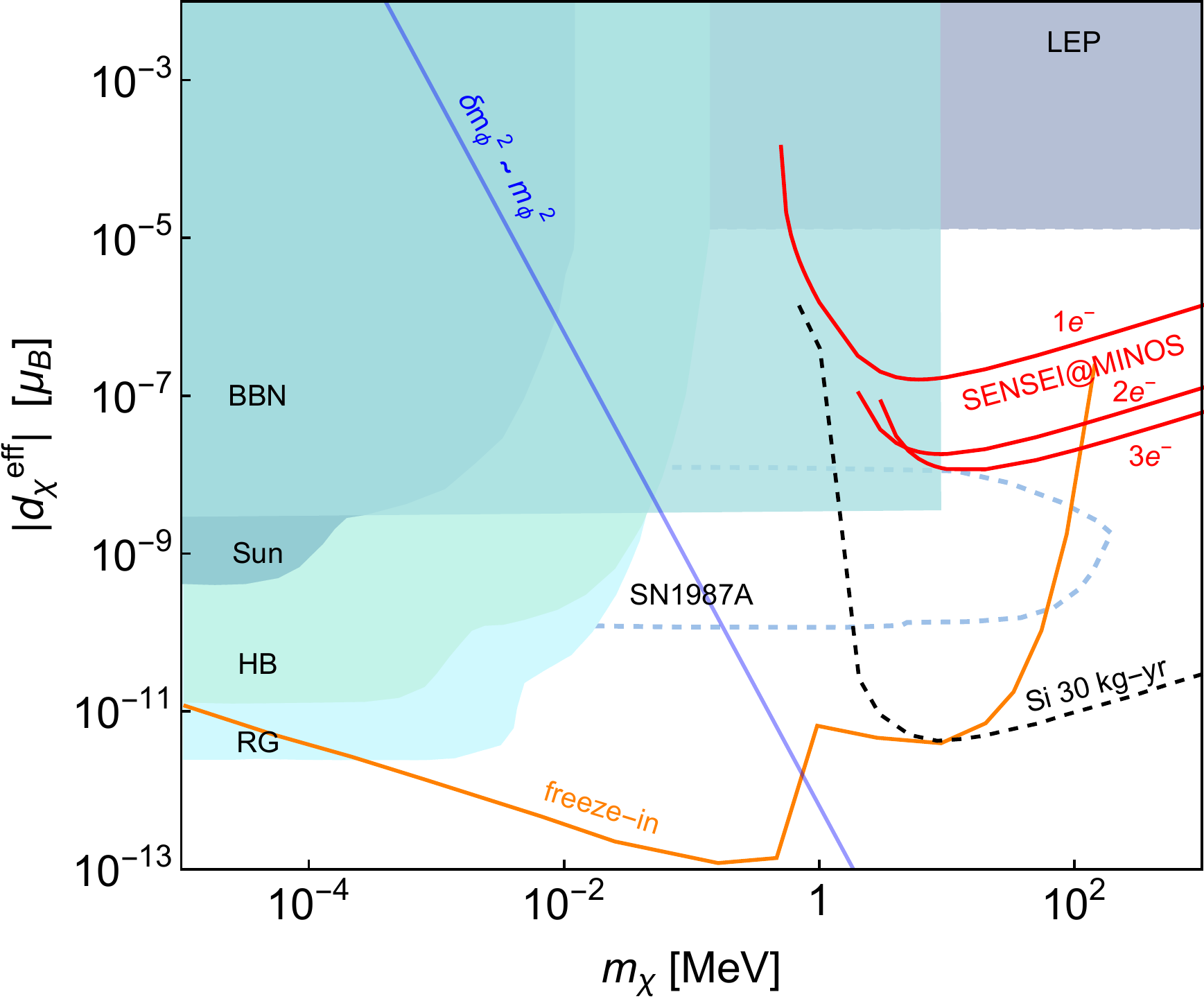}
\end{center}
\caption{Constraints on a constant EDM signal from: stellar cooling of red-giant stars (RG), horizontal branch stars (HB), and the Sun; cooling of the proto-neutron star of SN1987A; Big Bang nucleosynthesis (BBN), assuming a reheating temperature of $10~\mathrm{MeV}$~\cite{Chang:2019xva}; collider constraints using the mono-photon channel (LEP)~\cite{Chu:2018qrm}; and recent direct detection constraints from the SENSEI collaboration, showing the constraint coming from each individual electron bin~\cite{2004_11378}, along with the projected sensitivity of a silicon 30 kg-yr experiment~\cite{Chang:2019xva}. Analogous constraints and projections could be added for the XENON experiments, SuperCDMS, and DAMIC. Also shown is the line defined by Eq.~(\ref{eq:finetuningconst}), above which our effective field theory description likely requires a degree of fine-tuning to keep the dark axion light enough, as well as a line that indicates where freeze-in provides the full DM relic abundance, again assuming a reheating temperature of  $10~\mathrm{MeV}$~\cite{Chang:2019xva}.}
\label{fig:EDMconstraints}
\end{figure*}

In the absence of any undulation, the existence of a DM EDM is constrained by various processes in astrophysics, cosmology, and colliders,  as well as by the DM direct detection experiments. We have indicated these constraints in Fig.~\ref{fig:EDMconstraints}, which is based on Refs.~\cite{1908_00553,Chang:2019xva}. The astrophysical constraints plotted include stellar cooling of red giant stars (labelled `RG' in Fig.~\ref{fig:EDMconstraints}), horizontal branch stars (`HB'), and the Sun, all calculated in~\cite{1908_00553}, as well as a tentative limit coming from the cooling rate of SN1987A.\footnote{The form of this constraint, calculated in~\cite{1908_00553} (following~\cite{Chu:2018qrm}), is highly dependent on the modelling of the supernova explosion, assuming in particular that it is neutrino-driven. Other reasonable models give no bounds on the DM EDM~\cite{Blum:2016afe,Bar:2019ifz}. } On the cosmological side we plot a bound from Big Bang nucleosynthesis (`BBN'), which arises from the fact that light DM can come into equilibrium with SM particles in the early Universe, thus affecting its expansion. This depends on the reheating temperature which we take to be $10~\mathrm{MeV}$ (taking a higher reheating temperature will give a more stringent bound)~\cite{Chang:2019xva}. We have also included a line (orange) that indicates where the freeze-in mechanism~\cite{McDonald:2001vt,Hall:2009bx} accounts for the observed DM relic abundance, at the same reheating temperature~\cite{Chang:2019xva}. Finally, there are collider physics constraints which are flat in $m_\chi$ for sub-GeV dark matter masses, the strongest of which comes from the mono-photon channel at LEP~\cite{Achard:2003tx}, giving a constraint~\cite{Fortin:2011hv,Chu:2018qrm} plotted in Fig.~\ref{fig:EDMconstraints}.

Our primary interest, however, lies in the constraints coming from direct detection experiments. As an example, in Fig.~\ref{fig:EDMconstraints} we have plotted constraints (labeled SENSEI@MINOS) based on data recently released by the SENSEI collaboration~\cite{2004_11378}, obtained using a prototype Skipper-Charge-Coupled-Device (Skipper-CCD),\footnote{At the time of writing, these constraints are currently world-leading for DM masses $m_\chi \in [0.6,5]$ MeV, as well as in the low mass window $m_\chi \in [1.2,12.8]$ eV.} as well as a future projection for a silicon 30 kg-year experiment (which we have taken from Ref.~\cite{Chang:2019xva}). For reasons that will soon become clear, we plot individual curves corresponding to the constraints from the SENSEI data~\cite{2004_11378} in each electron bin, which we calculate using \texttt{QEdark}~\cite{1108_5383,1509_01598}.
Here, the reference DM-electron scattering cross-section $\overline{\sigma}_e$ is computed as $\overline{\sigma}_e=4(d^\text{eff}_\chi)^2\mu_{\chi e}^2/\alpha m_e^2$  following Ref.~\cite{Chang:2019xva}, where $\alpha$ is the fine-structure constant and $\mu_{\chi e}$ is the reduced mass between the electron and $\chi$.\footnote{For these calculations we assume the local dark matter density is $\rho_{DM}=0.3\mathrm{GeV}/\mathrm{cm}^{3}$, the dark matter escape velocity is $600 \mathrm{km}/\mathrm{s}$, the mean local velocity of dark matter is $230\mathrm{km}/\mathrm{s}$ and the average Earth velocity is $240\mathrm{km}/\mathrm{s}$.} 
The data from the $1e^-$ bin is significantly less constraining than the multi-electron bins because the background rate is high, with the $\geq 3e^-$ bin being essentially background-free. On the other hand, due to kinematic thresholds, the multi-electron bins lose sensitivity for dark matter masses lighter than 1 MeV or so. 

We chose to plot the SENSEI constraints in Fig.~\ref{fig:EDMconstraints} due to the ease with which they could be used to highlight the scaling of the constraints with respect to the background in each bin. Note that analogous constraints could be added to Fig.~\ref{fig:EDMconstraints} for the XENON experiments~\cite{Cao:2014jsa,Aprile:2017aty} (see \emph{e.g.}~\cite{Essig:2012yx}), the SuperCDMS experiment~\cite{Agnese:2016cpb,Agnese:2018col} and the DAMIC experiment~\cite{PhysRevLett.123.181802}.

\subsection*{Detecting an undulating signal}

We now turn to the prospect of detecting a signal with an undulating component. In order to extract limits on such a component from the real direct detection data of the XENON, SENSEI, SuperCDMS, and DAMIC collaborations, one requires more detailed information concerning the data acquisition (in particular, its binning in time) than is publicly available, and so we are not in a position to venture realistic limits in this paper. 

\begin{figure}[t]
\begin{center}
\includegraphics[width=3in]{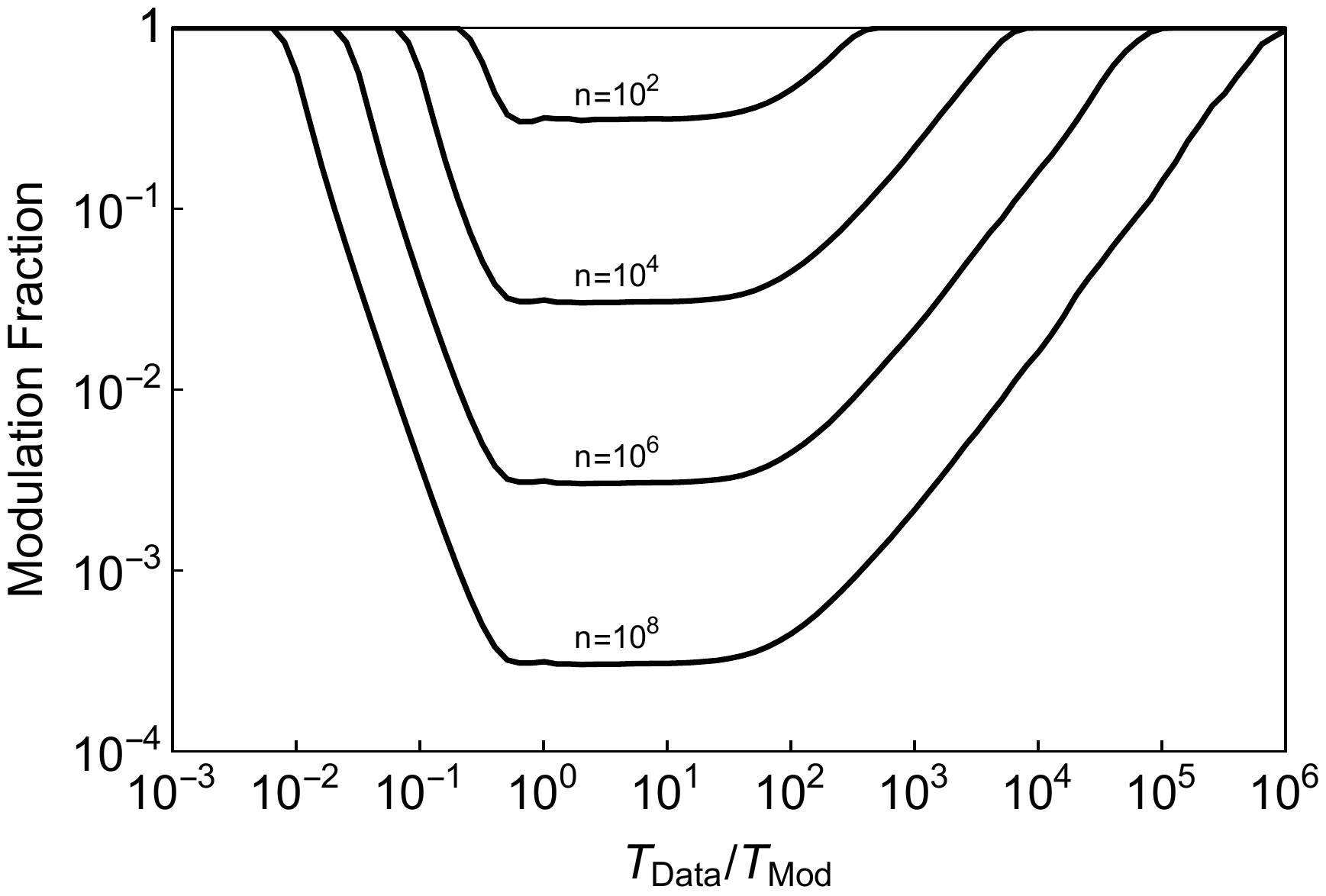}
\end{center}
\caption{Exclusion power for a modulation fraction for $n$ total events distributed evenly over a time interval with $365$ bins.  When the frequency is greater than the inverse of the total data taking period, $T_{\text{Data}}/T_{\text{Mod}} \sim 1$, and smaller than the inverse bin width, $T_{\text{Data}}/T_{\text{Mod}} \sim 365$, there is strong additional sensitivity to any modulations present, reflected in strong expected exclusion limits, with the enhanced sensitivity improving statistically as $1/\sqrt{n}$.}
\label{fig:modsens}
\end{figure}

Instead, to get an idea of the exclusion or discovery power that an undulating signal offers, we consider a pseudo-experiment in which we `read out' data from a flat distribution in $365$ randomly chosen bins over a period $T_{\text{Data}}$.  The time period could be any timescale relevant to a detection strategy.  To estimate the sensitivity to oscillations we consider the sensitivity to the distribution $R(t) \propto q_1+q_2\sin(2 \pi \nu t)$, which we assume models the combined background plus potential signal rate one could observe in a detector.  We emphasise that the constant component $q_1$ may contain both signal and background contributions, which we might not be able to separate, while the undulating component $q_2$ is assumed to be pure signal.  This distribution is characterized by two parameters, the frequency $\nu= 1/T_{\text{Mod}}$ and the undulation fraction $q_1/q_2$.  Then, given such a pseudo-dataset which is not oscillating, one can ask what undulation fraction $q_1/q_2$ would be excluded at a given confidence level.  In Fig.~\ref{fig:modsens} we plot the corresponding 10\% exclusion contours one would find for $q_1/q_2$ if a flat distribution were observed, where the label on each contour indicates the total number of signal plus background events distributed over the entire readout.

When the background component is small, or known precisely, the search for an unexpected undulation could be used to further characterise any dark matter discovery which has arisen through standard searches for an excess of events above background.  However, the real strength of a search for oscillations is found in experimental regions where the background is large or poorly known.  Backgrounds of this sort hamper the standard search for dark matter since the sensitivity can only be as strong as the systematic uncertainties on the background estimate allow.    This is the case for the SENSEI@MINOS lines in Fig.~\ref{fig:EDMconstraints}, where one can see that the strongest search limits are for di- and tri-electron events due to the very low background rates.  As one goes to lower masses, however, one is inevitably pushed towards relying on single electron events. It is precisely for this mass range that searches for undulating signals could enhance discovery potential, by factors comparable to the enhancement factors shown in Fig.~\ref{fig:modsens}.  As light dark matter searches progress to lower cross sections the interplay between signal and background will evolve.  If at lower cross sections additional backgrounds arise, or if they become more difficult to estimate at a level commensurate with the integrated exposure, then the enhanced sensitivity to undulating signatures will become of greater utility.

Of course, if there are modulating backgrounds then they must be accounted for.  They could be vetoed by omitting frequency ranges corresponding to experimental effects relating to power sources, or terrestrial effects, with the efficacy of the veto depending on how well the background is understood.  In this vein, the discovery of an unexpected modulating signal may reveal previously unknown background sources which may aid in background subtraction.  


\subsection*{Implications for the microscopic theory}

We now discuss whether the regions of parameter space probed by these direct detection contours can be realistically populated by the dark axion-induced UEDM couplings described in \S \ref{sec:EFT}, given the consistency bound (\ref{eq:EFTbound}).
We then speculate on what this might imply about the underlying microscopic theory, anchoring our (limited) discussion in the three scenarios sketched in Appendix \ref{app}; namely, that $\chi$ is (i) a strict QCD-like dark baryon, (ii) a large-$N_c$ dark baryon, or (iii) a fundamental fermion with perturbatively generated EDM couplings. 

Recall that the dark axion mass $m_\phi$ is identified with the frequency of the undulating signal. For the upcoming SENSEI experiment, one would expect sensitivity to undulation frequencies of order $m_\phi \sim (1~\text{week})^{-1}$ or slower,\footnote{In the recent preliminary study of the silicon Skipper-CCDs to be used in the SENSEI instrument~\cite{2004_11378}, the Skipper-CCDs were exposed for 20 hour periods for each read out. } while SuperCDMS would likely be sensitive to faster undulation frequencies thanks to more frequent data readout, say of order $m_\phi \sim (1~\text{min})^{-1}$. For the case of a strictly QCD-like dark sector, as shown in Appendix~\ref{app} we expect the parametric dependence
\be
d_\chi \approx 4\pi^2 e \frac{m_\phi^2 f^2}{m_\chi^5}~~,
\ee
which would require severely trans-Planckian values of the symmetry breaking scale $f$ in order to generate large enough Wilson coefficients $d_\chi$ for $|d^\text{eff}_\chi|$ to be realistically detectable by SENSEI or SuperCDMS, even with the aid of an undulating signal.

Arguably, this seems to point towards UV models for the dark sector which deviate from the strict QCD analogy. 
Firstly, if $\chi$ were a dark baryon for a QCD-like dark sector but with a large number of colours $N_c$, the Wilson coefficient $d_\chi$ would be expected to receive a large-$N_c$ enhancement. The SENSEI experiment could then be sensitive to UEDM signals with $f$ brought down to the Planck scale, but only for an exponentially  large number of colours,\footnote{While a huge `number of dark colours' might sound ludicrous, one could instead employ a holographic perspective \cite{Maldacena:1997re} and loosely interpret this large-$N_c$ limit rather as an indication of a different description of the microscopic theory that enjoys nearly-conformal dynamics.} and notwithstanding the problem that, if there is an MeV scale dark baryon, it would likely be accompanied by charged baryons at similar masses which would have been observed.

Much more palatable, then, is the scenario that $\chi$ is simply an elementary dark-charged fermion. Then the Wilson coefficient $d_\chi$ is no longer necessarily tied to the mass of the dark axion. We find that the interesting regions of parameter space in Fig.~\ref{fig:EDMconstraints} can be probed for safely sub-Planckian values of $f$. For more theoretical details, we refer the reader to Appendix \ref{app}.


\subsection*{Detecting the dark axion}

The dark axion $\phi$, whose oscillations are responsible for the undulating EDM of the fermionic DM $\chi$, is itself cosmologically abundant, constituting a fraction $r \lesssim 0.1$ of the local DM energy density. One might therefore hope to probe the UEDM scenario that we have proposed by hunting for the dark axion.

Recall that the mass $m_\phi$ of the dark axion sets the undulation frequency $\nu=m_\phi/2\pi$ of the UEDM and thus of the DM scattering rate of electrons, discussed in the preceeding subsections. Thus, if an undulation scattering rate were observed in a direct detection experiment, our model would make a striking prediction of the precise mass required of the dark axion. 

As mentioned above, in principle $m_\phi$ is a free parameter of our theory. However, our greatest interest lies in masses in the ballpark range of $10^{-20} \text{~eV} \lesssim m_\phi \lesssim 10^{-15}$ eV or so, for which present and future DM-electron scattering experiments (including SENSEI, SuperCDMS, XENON1T, and DAMIC) could be sensitive to the undulation. While this is an exceptionally low mass range, a recent proposal to detect ALPs through superconducting resonant photon frequency conversion~\cite{Berlin:2019ahk} (see also~\cite{Kahn:2016aff}) expects sensitivity to ALP masses as low as $10^{-14}$ eV, for realistic experiment run-times. 
Moreover, exploiting a broadband pump mode spectrum could bring sensitivity to ALP couplings for lower masses still (for realistic run-times)~\cite{berlin:future}, likely reaching our range of primary interest.


\section{Summary}
\label{sec:sum}
The laws of nature are not $CP$-symmetric, and $CP$-violation was discovered through the time-dependent oscillations of neutral kaons.  Indeed, oscillations of some form or another played the lead role in the physics of $CP$-violation until the 1990s, when direct $CP$-violation was finally observed.  There is no reason to expect that the dark sector, nor dark sector interactions with the visible sector, should not feature $CP$-violation in some form.  The marriage of these two elements is suggestive that the consideration of $CP$-violation and the associated possibility of time-dependent dark undulations should feature in any broad programme to uncover the dark sector.

Motivated by this, using a simple model involving a neutral stable dark fermion and a $CP$-odd axion-like particle which sources the EDM of the fermion, we have demonstrated that time-dependent \emph{coupling} undulations could lead to unique smoking gun signatures in direct detection experiments.  Since these undulations do not have their origin in celestial dynamics, the time scale is unrelated to celestial timescales (in this model being set by the mass of the axion-like particle).  As this is a free parameter, searches for oscillations in essentially any frequency range are warranted.

From the theoretical point of view, we briefly consider three kinds of microscopic dynamics that might underlie the effective description of DM interacting via an UEDM. Of the three, the scenario in which the DM is an elementary fermion that acquires its UEDM from couplings to heavy charged matter appears the most realistic, in terms of being probed by experiments in the near future.

By its nature, to observe an undulation requires a large number of signal events. Thus, for direct detection experiments which feature low calculable backgrounds, undulations will never serve as a discovery mode (although they could play a key role in the subsequent stages of characterising the nature of the dark sector).  On the other hand, in regions where the backgrounds are large and have systematic uncertainties which are difficult to reduce, as in very low mass DM-electron scattering which results in only single electron events, then large numbers of DM scattering events could hide below large backgrounds.  In this instance, searches for undulations could provide the strongest statistical evidence for new physics, potentially enabling discovery when canonical time-independent analyses cannot disentangle signal from background.  Pragmatically, this is arguably the best motivation to search for undulating DM.

Recently XENON1T has reported an excess in just such a region~\cite{Aprile:2020tmw}.  While this excess most likely has it's origins in more mundane effects, it is interesting to speculate if it could be the harbinger of a DM discovery.  For DM scattering, rather than absorption, vanilla scenarios are disfavoured due to the observed energy spectrum of the excess.  However, exotic DM distribution subcomponents with high velocity components~\cite{Kannike:2020agf,Chen:2020gcl,DelleRose:2020pbh,Ko:2020gdg}, exothermic DM scattering~\cite{Su:2020zny,Harigaya:2020ckz,Lee:2020wmh,Baryakhtar:2020rwy,Bloch:2020uzh,An:2020tcg}, or scattering which produces photons \cite{Bell:2020bes,Paz:2020pbc}, can provide possible DM scattering explanations.  In these cases, if the DM coupling undulates then, irrespective of kinematic details, the scattering rate will undulate, although at present with low statistics the events are consistent with both constant and annual modulations.  Nonetheless, in future a search for time-dependent signatures such as oscillations are warranted.

A lesson learned from nascent ventures into light DM direct detection is that existing approaches, and even apparatus, can often be repurposed to search for DM in previously unexplored kinematic regimes.  This demonstrates the extraordinary versatility of technologies and detection strategies which have been developed over many years.  Furthermore, these relatively recent developments highlight the importance of theoretical and experimental perspectives which evolve rapidly in response to advances on either side.  This work demonstrates that a detection effort which also includes searches for DM undulations, in addition to annual and diurnal modulations, to directional signatures, and to standard constant isotropic scattering, could enhance the discovery and/or signal characterisation power of light DM direct detection efforts.


\bigskip\bigskip

\acknowledgments{
We are grateful to Rouven Essig and Tien-Tien Yu for help with \texttt{QEdark}, and to the Cambridge pheno working group, Malcolm Fairbairn, and Raffaele  Tito  D’Agnolo for discussions.  We are particularly grateful to Simon Knapen for numerous discussions through the later development of this work and to Simon Knapen and Tien-Tien Yu for detailed comments on this draft. This work has been  supported by STFC consolidated grants ST/P000681/1 and ST/S505316/1.
}

\appendix


\section{Three microscopic candidates}
\label{app}


In this Appendix, we give more details concerning three possible microscopic theories that could match onto the undulating DM effective field theory, commenting in particular on the expected size of the EFT Wilson coefficient $d_\chi = ce/\Lambda$ in each scenario.

\medskip

\underline{1. QCD-like dark baryon}

\medskip
Firstly, consider a scenario in which the dark sector is very closely analogous to QCD plus an axion, {\em i.e.} an $SU(3)$ gauge theory at strong coupling. In that case, the quantity $d_\chi \equiv ce/\Lambda$ is precisely analogous to the EDM of the neutron in QCD, in units of the effective $\theta$ angle (see {\em e.g.}~\cite{Pospelov:2005pr} for a review). We would expect
\be
d_\chi \approx \frac{e}{m_\chi^2} \frac{m^\text{dark}_q}{2}~~,
\ee
where $m^\text{dark}_q$ is a typical mass for the fundamental dark quarks. Note that $d_\chi$ scales linearly with $m^\text{dark}_q$, because in the limit that $m^\text{dark}_q$ vanishes one can rotate away the effective theta angle responsible for the dark EDM by a chiral transformation. Importantly, the dark axion mass squared is also expected to scale linearly with $m^\text{dark}_q$, assuming that the explicit breaking of the PQ-like $U(1)_A$ symmetry is dominated by the chiral anomaly contribution:
\be
m_\phi^2 \approx \frac{m^\text{dark}_q |\langle \overline{q} q \rangle_{\text{d}}|}{f^2}~~,
\ee
where $\langle \overline{q} q \rangle_{\text{d}}$ is the quark condensate that spontaneously breaks chiral symmetry in the QCD-like dark sector.
Supposing moreover that the Dirac fermion $\chi$ is, like the neutron of QCD, a composite state that is bound due to the strong dark dynamics, then its masss is bound to the quark condensate, $m_\chi^3 \approx 8 \pi^2 \langle \overline{q} q \rangle_{\text{d}}$ \cite{Ioffe:1981kw}. The upshot of these relations is that the dark matter EDM coefficient scales like
\be \label{eq:app_QCD}
d_\chi \approx 4\pi^2 e \frac{m_\phi^2 f^2}{m_\chi^5}~~,
\ee
varying with the fifth inverse power of $m_\chi$. 

Note that in this scenario one would invariably expect charged baryons.  As a result, it is only a viable possibility for DM with mass at or above the electroweak scale, and not for DM masses in the MeV r\'egime.

\medskip

\underline{2. Large-$N_c$ dark baryon}

\medskip

There is no good reason, however, to suppose that the dynamics in the dark sector should be so strictly analogous to real-world QCD. A simple variant to consider is a QCD-like theory with gauge group $SU(N_c)$, but with a large number of colours $N_c$. In this scenario, which is under better theoretical control than QCD thanks to the $1/N_c$ expansion, the dark matter EDM is expected to receive a large-$N_c$ enhancement by a factor $\sim N_c \ln N_c$~\cite{Crewther:1979pi,Riggs:1992jh}.  Again, in this scenario one would invariably expect charged baryons, thus it is also only a viable possibility for DM with mass at or above the electroweak scale.

\medskip

\underline{3. Dark elementary fermion}

\medskip

As a third (and simpler) possibility, one could imagine that the dark fermion $\chi$ is not a composite state at all, but is rather an elementary SM singlet fermion. The EDM-like couplings in Eq. (\ref{eq:lagrangian}) could then be generated, for example, by integrating out heavy charged particles (such as a fermion and a scalar, following Refs.~\cite{Graham:2012su,Chang:2019xva}), with the factor of $\sin(\phi/f)$ in (\ref{eq:lagrangian}) resulting from the $CP$ structure of the EDM interaction. Unlike in the previous two scenarios, this does not mandate the existence of any further charged states, and so the DM mass $m_\chi$ can safely reside in the MeV range of most interest to the electron-scattering detection experiments discussed in the main text. 

In this case, one expects the Wilson coefficient to be~\cite{Graham:2012su}
\be \label{eq:app_elementary}
d_\chi \approx \frac{eg^2}{8\pi^2 M}~~,
\ee
if both the charged scalar and fermion have mass $M$ and couplings $g$ to the dark matter particle $\chi$. We emphasize that $d_\chi$ is essentially decoupled from the dark axion mass $m_\phi$, provided only that the EFT bound (\ref{eq:EFTbound}) is satisfied.
Thus,  $d_\chi$ and $m_\phi$ can, in this scenario, be treated as almost independent phenomenological parameters that map onto the magnitude and the frequency of the UEDM. There is, however, now an additional `naturalness bound' coming from the expected correction to the mass of $\chi$ in the presence of the heavy charged states to which it couples. Based on the cutoff dependence of the appropriate one-loop diagrams, one expects~\cite{Graham:2012su,Chang:2019xva}
\be
\delta m_\chi \sim \frac{M^2}{2e} d_\chi \lesssim m_\chi,
\ee
resulting on an upper limit on $d_\chi$ that scales linearly with $m_\chi$. The dependence on $M$ means that the naturalness bound admits larger UEDMs for lighter $M$. Charged scalars as light as 100 GeV are still compatible with collider bounds~\cite{Egana-Ugrinovic:2018roi,PhysRevD.98.030001}, in which case there is a region satisfying both naturalness bounds that will be probed by, say, a silicon 30 kg-year direct detection experiment (see Fig.~\ref{fig:EDMconstraints}).

\bibliographystyle{JHEP}
\bibliography{references}

\end{document}